\documentclass{aa}
\usepackage{graphicx}

\begin{document}


\title{
Catalogue of cataclysmic binaries, low--mass X--ray binaries
and related objects (Seventh edition)}

\author{Hans Ritter$^{1}$ and Ulrich Kolb$^{1,2}$} 

\institute{$^1$Max-Planck-Institut f\"{u}r Astrophysik,
	       Karl-Schwarzschild-Str.~1,
               D-85741 Garching, Germany \\ 
           $^2$Department of Physics and Astronomy, The Open University, 
               Milton Keynes MK7 6AA, U.K.
}
\offprints{H.~Ritter, \email{hsr@mpa-garching.mpg.de}}

\authorrunning{Hans Ritter and Ulrich Kolb}
\titlerunning{Catalogue of CVs, LMXBs and related objects}

\date{Received ; accepted}


\abstract{
The catalogue lists coordinates, apparent magnitudes, orbital
parameters, and stellar parameters of the components and other
characteristc properties of 472 cataclysmic binaries, 71 low--mass
X--ray binaries and 113 related objects with known or suspected
orbital periods together with a comprehensive selection of the
relevant recent literature. In addition the catalogue contains a
list of references to published finding charts for 635 of the 656 
objects, and a cross--reference list of alias object designations. 
Literature published before 1 January 2003 has, as far as possible,
been taken into account. All data can be accessed via the dedicated
catalogue webpage at {\tt http://www.mpa-garching.mpg.de/RKcat/} and
{\tt http://physics.open.ac.uk/RKcat/}. We will update the information 
given on the catalogue webpage regularly, initially every six months. 

}
 
\maketitle

\section{Introduction to the 7th Edition}

Five and a half years after the publication of the previous (6th) 
edition of the {\em Catalogue of Cataclysmic Binaries, Low--Mass X--Ray 
Binaries and Related Objects} (Ritter \& Kolb 1998, hereafter RK98)),
the amount of new  literature  and the number of new objects to be
included have again grown so much that it seems worthwhile to publish
an updated edition. The philosophy  and the  purpose of this catalogue
(now in its 7th edition) have been outlined in the preface to the 3rd
edition (Ritter 1984, hereafter R84) and will not be repeated
here. Rather we briefly recall some of the developments which,
over the past five and a half years, have had (and still have) a major
impact on this catalogue:
\begin{enumerate}
\item The online and living edition of
      {\em A Catalog and Atlas of Cataclysmic Variables} by Downes et
      al. (2001, hereafter referred to as DWSRKD) is now the primary 
      source for accurate coordinates and finding charts for
      cataclysmic variables. The DWSRKD catalogue supersedes earlier
      editions by Downes \& Shara (1993) and  Downes, Webbink 
      \& Shara (1997), includes an updated version of {\em A
      Reference Catalogue and Atlas of Galactic Novae} by Duerbeck
      (1987), and is supplemented by orbital period information
      by the present authors.
\item An increasing number of CVs are found
      serendipitously from large surveys such as e.g.\ the Slone Digital
      Sky Survey (see e.g. Szkody et al. 2002), or from The All Sky 
      Automated Survey at
      \verb+http://www.astrouw.edu.pl/~gp/asas/asas.html+. 
\item As a result of systematic searches for
      potential double degenerate Type Ia supernova progenitors (see
      e.g. Napiwotzki et al. 2001; Marsh 2000; Morales-Rueda et al.
      2002) the number of detached short-period double white dwarf
      systems has increased dramatically and is likely to further
      increase substantially in the future.   
\item We acknowledge the continuing contributions of amateur 
      astronomers to the field of cataclysmic variables, in particular
      to tracking down the superhump periods of SU~UMa stars. Results
      of such amateur activities can e.g. be found on the webpages of
      the Variable Star Network (VSNET) at
      {\verb+http://vsnet.kusastro.kyoto-u.ac.jp/vsnet/+} and of the
      Center for Backyard Astrophysics (CBA) at
      {\verb+http://cba.phys.columbia.edu/+}.   
\item Due to the much increased sensitivity and
      spatial resolution achievable with present--day X--ray satellites
      such as {\em Chandra} or {\em XMM-Newton}, as well as with large
      ground--based optical telescopes (e.g. the {\em VLT} or the {\em
      Keck} telescopes) the number of cataclysmic variables (CVs) and
      low--mass X--ray binaries (LMXBs) found in globular clusters and
      in external galaxies (M31, M51) is also rapidly increasing. 
\item G\"ansicke \& Kube (2002) are in the process of establishing
      {\bf CVcat}, a data base on CVs that aims at a comprehensive
      collection of all available data on CVs, and to make them
      electronically available to the community. Thus its aims go way
      beyond those of DWSRKD or the present catalogue. 
\end{enumerate}

In view of the activities by G\"ansicke \& Kube (2002) and the fact
that with DWSRKD there is already an online CV catalogue one might ask
whether publishing yet another one is justified. 
DWSRKD and this catalogue have existed in parallel in paper form
before going electronic, and are complementary as far as CVs are 
concerned. Our catalogue focusses on data related to the objects'
binary nature, while DWSRKD gives data that are useful for
observational programmes. In addition, our catalogue includes LMXBs
and related objects, for which there is no comparable alternative
source (the most recent catalogue for LMXBs is by Liu, van Paradijs
\& van den Heuvel 2001). 
The CVcat project by G\"ansicke \& Kube (2002), on the other hand,
is still under development, and it remains to be seen whether this
project will render our catalogue superfluous.  

Compared with the 6th edition, the number of objects listed has
increased by almost $60 \%$. Accordingly the current version of
this catalogue provides tabulated data and references for 656 objects
(472 cataclysmic binaries, 71 low--mass X--ray binaries and 113
related objects).

No attempt has been made to provide a complete bibliography. Rather,
the aim is to give a selection of the most recent and most relevant
references that should allow the user to navigate through the
literature addressing mainly the binary properties of the objects in
question. References that were already included in RK98 are only
repeated if they are required  for cross--reference (see below).  
Yet more references can be found in the previous four editions (R84,
Ritter 1987, hereafter R87, R90, and RK98) of the catalogue.
Accordingly the 7th edition provides:       
\begin{itemize}
\item the tables for all three object classes  (cataclysmic 
      binaries, low--mass X--ray binaries and related objects) in full;
\item for each catalogued object a selection of references to the 
      literature, published after 30 June 1997 (the deadline of RK98).
      Earlier references are only included if they are needed for
      cross--reference, or in cases where there have been few or no
      new publications of relevance;    
\item a list of selected references to published finding charts. 
      Additional references may be found in previous editions;                
\item the {\em Who's Who?} file, a cross--reference list of alias names of
      the objects catalogued. 
\end {itemize}
Thus the catalogue is complete and self--contained in the tables  
and in giving cross--references to alternative object designations. 

Every effort has been made to avoid errors and to keep the lists up to
date. Nevertheless, the authors are well aware of the fact that also 
this edition will contain errors and may be incomplete with regard to 
the criteria stated in the preface to the 3rd edition (R84). It is 
certainly  incomplete with  respect to the references quoted. 

All the tabular material contained in this catalogue is published in
electronic form only. It is available from the dedicated catalogue
webpages at the Max--Planck--Institut f\"ur Astrophysik (MPA)  
\verb+http://www.mpa-garching.mpg.de/RKcat/+, and the Open University
(OU) \verb+http://physics.open.ac.uk/RKcat/+.

In addition to the electronic version we provide postscript files for
a printable stand alone version at both websites.

For the current version of this catalogue, literature published before
1 January 2003 has, as far as possible, been taken into account.
This version is also accessible from the CDS at
\verb+http://cdsweb.u-strasbg.fr/Abstract.html+.

For the future it is our intention to provide semi--annual updates of
the catalogue on the dedicated catalogue webpages.

\section{Description of the catalogue}

The objects listed in this catalogue are subdivided into three main
object classes, i.e. into {\bf Cataclysmic Binaries}, {\bf Low--Mass  
X--Ray Binaries} and {\bf Related Objects}. The defining 
characteristics of the three object classes used here are the 
following:                                    

\begin{description}
\item[{Cataclysmic  Binaries}]
are semi--detached binaries consisting of a white dwarf primary (or   
a white dwarf precursor) and a low--mass secondary which is filling   
its critical Roche lobe. The secondary is not necessarily unevolved.
It may even be a highly evolved star as for example in the case of
the AM CVn--type stars. A more detailed description of the main 
characteristics of these objects may be found in Warner (1995) or
Hellier (2001).
                                                                        
In addition, we list among the cataclysmic binaries also the 
supersoft binary X--ray sources, because these too are semi--detached 
binaries containing a white dwarf, though one in a state of sustained 
nuclear burning. More information about these objects can be found in 
Greiner (1996, 2000).

\item[Low--Mass  X--Ray Binaries]
are  semi--detached binaries consisting of either a neutron star or
a black hole primary, and a low--mass secondary which is filling its
critical Roche lobe. Observationally they are distinguished from 
the luminous, massive X--ray binaries by the following main properties:
in general the spectra of the low--mass X-ray binaries (at maximum light)
are devoid of normal stellar absorption features. The ratio of their 
X--ray to optical luminosities is much larger than unity (typically it 
ranges from $\sim 10^2$ to $\sim 10^4$). A more detailed  description of 
the main characteristics of these objects may be found in the review 
articles in Lewin, van Paradijs \& van den Heuvel (1995).

\item[Related Objects]
are detached binaries consisting of either a white dwarf or a white 
dwarf precursor primary and of a low--mass secondary. The secondary 
may also be a highly evolved star. Further information may be found 
e.g. in Ritter (1986), Bond (1989), or de Kool \& Ritter (1993).
                                                                        
With one possible exception (HD~49798) we do not list among the 
related objects detached binaries containing a neutron star, or,  
for the lack of known objects, a black hole. Thus we explicitly 
exclude binary radio pulsars from our compilation because these are
documented elsewhere (e.g.\ in the Princeton pulsar catalogue 
(Taylor, Manchester \& Lyne 1993)) which is available online at 
{\verb+http://pulsar.princeton.edu/pulsar/+}
{\verb+catalog.shtml+}.
\end{description}

\bigskip

According to the subdivision into these three object classes the catalogue
consists of three major parts, hereafter referred to as {\em catalogue 
sections}. Each of the three catalogue sections is further subdivided into 
a {\em table section}, where a few characterizing parameters of the 
object are tabulated, and a {\em reference section}, where a 
selection of references is given. Within each of the table sections, the 
objects are listed in order of decreasing orbital period. In the 
corresponding reference section, however, the objects are listed in 
lexigraphical order.\par 

Limited information about where the values given in the tables are
taken from is provided as follows:  at the end of a reference from
which a given quantity, say {\verb+XYZ+}, was taken, this quantity is
given in parenthesis, i.e. as {\verb+(XYZ)+}. The  quantities for
which this is done are: the periods ({\verb+Orb.Per.+},
{\verb+2.  Per.+}, {\verb+3.  Per.+}, {\verb+4.  Per.+}), the spectral
types ({\verb+Spectr1+}, {\verb+Spectr2+), the mass ratio 
({\verb+M1/M2+}), the orbital inclination ({\verb+Incl+}), the masses 
({\verb+M1+}, {\verb+M2+}), and, where appropriate, the radii 
({\verb+R1+}, {\verb+R2+}) and the eccentricity ({\verb+e+}).

The catalogue is supplemented by a list giving references to published
{\em finding charts} of the objects. In this separate section, the
objects of all three classes are merged and listed in lexigraphical
order. The full form of abbreviated references used is given at the
end of this section.\par

Finally, the {\em Who's Who?} file contains a cross--reference list of
alias names of the objects catalogued. In order to keep this list
short, the full list of alternative object names appears only once for
each object and is to be found under the standard name used in this
catalogue. If an object is sought under one of its alternative names,
reference to the  standard name is given. Wherever possible the
variable name given in the 4th edition of the {\em General Catalogue
of Variable Stars} (Kholopov et al.\ 1985a, 1985b, 1987), or in
its online  version at 
{\verb+http://www.sai.msu.su/groups/cluster/gcvs/gcvs/+}, or in the 
{\em Name Lists of Variable Stars} (up to and including the 76th list
(Kazarovets, Samus \& Durlevich 2001)) is used as the standard name
here. This section includes also a list of references to various
catalogue acronyms that appear in this compilation. More complete
information of this kind may be found in {\em The First Dictionary of the 
Nomenclature of Celestial Objects} by Fernandez, Lortet \& Spite (1983),
its supplements Lortet (1986a, 1989b), and Lortet \& Spite (1986), in
the {\em Second Reference Dictionary of the Nomenclature of Celestial
Objects} by Lortet, Borde \& Ochsenbein (1994), or online via the
{\em Centre de donn\'ees astronomiques de Strasbourg (CDS)} at
{\verb+http://vizier.u-strasbg.fr/cgi-bin/Dic+}, or from the 
{\em Astronomical  Data Center (ADC)} at
{\verb+http://adc.gsfc.nasa.gov/adc/+}.

\acknowledgements
We wish to thank H.--C.~Thomas and V. Burwitz for keeping us informed
about the latest results regarding the optical identification and
follow--up observations of new CVs from the ROSAT All Sky Survey. We
also thank J. Thorstensen for providing us with information on
numerous new CVs with newly measured periods prior to pblication.
U.K. thanks the Max--Planck--Institut f\"ur Astrophysik for its
hospitality. This research has made use of the SIMBAD database,
operated at CDS, Strasbourg, France

\end{document}